\documentclass[10pt,a4paper]{article}
\usepackage{times,epsf,amsmath,amssymb,graphicx}
\newcommand{\ket}[1]{\left|\mathrm{#1}\right\rangle}
\newcommand{\dket}[1]{\left|\mathrm{#1}\right\rangle\rangle}

\newcommand{\dbra}[1]{\left\langle\langle\mathrm{#1}\right|}
\newcommand{\mean}[1]{\left\langle\mathrm{#1}\right\rangle}
\newcommand{\aver}[2]{\left\langle#1\left|#2\right|#1\right\rangle}

\title{Photon-number entangled states generation model with stimulated parametric down conversion}
\author{Oleksandr O. Gurin${}^1$, Vladyslav C. Usenko${}^2$, Constantin V. Usenko${}^1$\\
${}^1$ National Shevchenko University of Kyiv, \\Theoretical Physics Department, Kyiv, Ukraine\\
 ${}^2$ Bogolyubov Institute for Theoretical Physics \\of the National Academy of Sciences, Kyiv, Ukraine}
\date{}
\begin{document}
\maketitle

\begin{abstract}
We address the process of generation of the photon-number entangled states of light in the stimulated nonlinear parametric 
down conversion process and build the simple model describing the generation, not involving the traditional parametric approximation. 
The motion equations for the system of  pumping and two-mode outgoing field are solved for the case of the strong correlation between two modes, the evolution of the state parameters of the generated modes is obtained. The solution is briefly analyzed for particular types of photon-number entangled states.
\end{abstract}
\maketitle
\section{Introduction}
Nonclassical features of the states of light are of utmost interest within the past decades
as they became the basis for the quantum information theory and it's applications, in 
particular quantum computing, teleportation and cryptography \cite{qkd}. All of these are 
based on the fundamental principles of quantum mechanics and mainly involve the essential 
resource of quantum entanglement. The recent developments in the field are based on the 
continuous variables information coding \cite{cv} into the parameters of multi-level quantum 
systems as distinct from the discrete variables coding with a two-level system states. In 
the optical implementations it mainly corresponds to the consideration of multiphoton quantum 
states as the information carriers. At that the special attention is given to the entangled 
multiphoton states, in particular the photon-number entangled states of light (PNES), which provide 
with the strong correlation between the intensity measurements of the two spatial modes. 
Several communication protocols were proposed and partially implemented on the basis of the 
photon-number entangled states with the use of intensity difference \cite{funk,jp} or 
intensity fluctuations coding \cite{tmc}. Such states can be generally generated in the nonlinear process of 
parametric down conversion (PDC) \cite{klyshko,walls,mandel,perina} being the well 
known typical example of the quantum-mechanical event when the pumping photons 
excite the atoms of the nonlinear media, which emit photon pairs upon relaxation. While in 
the first experiments such pairs were observed independently, with the increase of process 
gain in the optical parametric amplifiers the generated multiphoton beams 
started to manifest the nonclassical features, in particular the strong correlation between 
modes (showing in the sub-shotnoise signal-idler intensity difference fluctuations) and 
became referred to as the twin beams \cite{twin1,twin2,twb_exp}. At that the modes marginal 
statistical distributions were thermal. The recent advances in the experimental technics 
(in particular, the use of the powerful short laser pulses for pumping and effective multiport 
photon-counters and CCD cameras) and usage of cavity-based parametric oscillators enabled 
several groups to obtain strongly correlated states 
of light which possess Poisson statistics \cite{haderka,fabre}. Meanwhile the traditional 
approach to description of the PDC process
is based on the parametric approximation, when the pumping mode is considered as classical \cite{walls}.
In the present paper we build a simple model of the PNES generation without the parametric approximation, thus keeping
nonlinearity in the operator equations. We make the statistical assumptions relevant to the considered states and obtain 
the solutions describing evolution of the state parameters. At that we assume that pumping is so intense 
that the atoms of the nonlinear media are excited coherently and the PDC process is stimulated. 
This work is the step towards the exact model describing the PDC-based generation of the PNES states.

\section{The photon-number entangled states}

The PNES may be written in the Fock number basis as 

\begin{equation}
\label{eq:pnes}
\dket{\Psi} = {\cal N}^{1/2}\:
\sum\limits_n {c_n \dket{n,n}} ,
\end{equation}
where $\dket {n,n}  = \left| n \right\rangle_1 
\otimes \left| n  \right\rangle _2$, indexes $1,2$ standing for the respective modes and ${\cal N}$ is a normalization factor. 

Their features are evident thereby: the amplitude mean values for each of the modes are zero, while the amplitudes are strongly correlated, i.e. in terms of the modes quantum operators $\mean{\hat{a}_1}=\mean{\hat{a}_2}=0$ and $\mean{\hat{a}_1\hat{a}_2} \equiv \lambda \neq 0$. These statistical properties of the states are used within the construction of the PNES generation model.

\section{Generation}
Let us describe the possible generation of PNES states by means of a PDC process. As it was already mentioned, 
we assume that the pumping pulses are intense enough to coherently excite the whole electron subsystem of 
the nonlinear media and so the pumping energy directly turns  to the energy of a generated two-mode electromagnetic 
field excitation. The system can be  described by the Hamiltonian:

\begin{equation}
\label{eq:hamil}
\hat{H}=\hbar\omega_1 \hat{a}^+_1\hat{a}_1+\hbar\omega_2 \hat{a}^+_2\hat{a} _2
+\hbar\omega_0 \hat{a}^+_0\hat{a}_0 + i\hbar\chi\left(\hat{a}^+_1 \hat{a}^+_2 \hat{a}_0-\hat{a}_1 \hat{a}_2 \hat{a}^+_0\right)
\end{equation}

which includes the interaction term, but doesn't include quantum operators standing for the electron subsystem excitation 
and relaxation and covers only the energy swap between the pumping and  emitted fields. We assume that the excitation and relaxation takes place  immediately, i.e. the model is adiabatic. 

The quantum operators for each of the two modes of the outgoing field satisfy the Heisenberg equations:

\begin{equation}
i\hbar\frac{d}{dt}\hat{a}_{1,2}=\left[\hat{a}_{1,2}\hat{H}\right]
=\hbar\omega_{1,2} \hat{a}_{1,2}+i\hbar\chi\hat{a}^+_{2,1} \hat{a}_0,
\end{equation}

similarly for the hermitian conjugate. The equations lead to ones for the operators 
$\hat{n}_{1,2}=\hat{a}^+_{1,2}\hat{a}_{1,2}$, which correspond to the photon numbers in each of the two generated modes:

\begin{equation}
i\hbar\frac{d}{dt}\hat{n}_{1}=
i\hbar\frac{d}{dt}\hat{n}_2
=i\hbar\chi\hat{a}_2 \hat{a}_1 \hat{a}^+_0
+i\hbar\chi\hat{a}^+_1\hat{a}^+_2 \hat{a}_0
\end{equation}

At that the difference between photon numbers in the two modes vanishes:

\begin{equation}
\frac{d}{dt}\left(\hat{n}_1-\hat{n}_2\right)=0
\end{equation}

which means that 

\begin{equation}
\mean{\hat{n}_1-\hat{n}_2}=0 
\end{equation}

if initially both of the modes aren't excited.

Introducing operators $\hat{A}=\hat{a}_1\hat{a}_2$, $\hat{A}^+=\hat{a}_1^+\hat{a}_2^+$ and $\hat{N}=\hat{n}_2 +\hat{n}_1$ 
we obtain the following system of equations:

\begin{equation}
\frac{d}{dt}\hat{A}=-i\omega \hat{A}+\chi\left(\hat{N}+1\right)\hat{a}_0
\end{equation}

\begin{equation}
\frac{d}{dt}\hat{A}^+=i\omega \hat{A}^++\chi\left(\hat{N}+1\right)\hat{a}_0^+
\end{equation}

\begin{equation}
\frac{d}{dt}\hat{N}=2\chi\left(\hat{A}\hat{a}_0^++\hat{A}^+\hat{a}_0\right)
\end{equation}

We suppose that pumping is coherent and the generated state is PNES, i.e. 

\(\mean{\hat{A}}=\dbra{\Psi}\hat{A}\dket{\Psi}=\lambda\left(t\right)\); \(\mean{\hat{a}_0}=\aver{\alpha}{\hat{a}_0}=\alpha\left(t\right)\);

\(\mean{\hat{A}^+}=\dbra{\Psi}\hat{A}^+\dket{\Psi}=\overline{\lambda}\left(t\right)\); \(\mean{\hat{a}_0^+}=\aver{\alpha}{\hat{a}_0^+}=\overline{\alpha}\left(t\right)\);

and that \(\mean{\hat{N}}=N\left(\lambda\left(t\right)\right) \equiv N\left(t\right)\).

Since we are describing the generated state as photon-number entangled, but not as the pair of coherent states, we may assume it's statistical independence from the pumping state but not the independence between the two modes. The overall system 
state can be presented as $\ket{\alpha}\otimes\dket {\Psi}$, 
while $\dket{\Psi}\neq\ket{\alpha_1}\otimes\ket{\alpha_2}$ which is the manifestation of modes' entanglement, i.e. $\mean{\hat{a}_1 \hat{a}_2 \hat{a}^+_0}=\mean{\hat{a}_1 \hat{a}_2 }\mean{\hat{a}^+_0}$, \(\mean{\hat{A}\hat{a}_0^+} =\lambda\left(t\right)\overline{\alpha}\left(t\right)\) and \(\mean{\hat{N}\hat{a}_0}=N\left(t\right)\alpha\left(t\right)\). Thus the system of equations for operators turns to the same for the state parameters:

\begin{equation}
\frac{d}{dt}\lambda\left(t\right)=-i\omega \lambda\left(t\right)+\chi\left(N\left(t\right) +1\right)\alpha\left(t\right)
\end{equation}

\begin{equation}
\frac{d}{dt}\overline{\lambda}\left(t\right)=i\omega \overline{\lambda}\left(t\right) +\chi\left(N\left(t\right)+1\right)\overline{\alpha}\left(t\right)
\end{equation}

\begin{equation}
\frac{d}{dt}N\left(t\right)=2\chi\left(\lambda\left(t\right)\overline{\alpha}\left(t\right) +\overline{\lambda}\left(t\right)\alpha\left(t\right)\right)
\end{equation}

Separating the slow amplitudes: \(\lambda\left(t\right)=e^{-i\omega t}\Lambda\left(t\right) \); 
\(\alpha\left(t\right)=e^{-i\omega t}a\left(t\right)\) and supposing that $a$ is real, thus $\Lambda$ is real, we obtain the system

\begin{equation}
\label{system.1}
\frac{d}{dt}\Lambda\left(t\right)=\chi\left(N\left(t\right)+1\right)a\left(t\right)
\end{equation}

\begin{equation}
\label{system.2}
\frac{d}{dt}N\left(t\right)=4\chi\Lambda\left(t\right)a\left(t\right)
\end{equation}

The solutions for the motion equations (\ref{system.1},\ref{system.2}) has the form

\begin{equation}
\label{sol}
\Lambda(t)=\sinh {\tau(t)} \cosh{\tau(t)};\ N=2 \sinh^2{\tau(t)},
\end{equation}

where

\begin{equation}
\label{tau}
\tau(t)=\chi \int_{-\infty}^t{a(t')dt'},
\end{equation}

assuming that initial conditions are \(t \to -\infty : \tau \to 0, N \to 0, \Lambda \to 0\).
The solutions describe the connection between the time profiles of the pumping and generated modes. The effect of the pumping profile is specified by the integral characteristic (\ref{tau}), in particular for the case of pulse pumping 
\(a\left(t\right)=\left\{t<0:0;t>T:0;0<t<T:a\right\}\): \(\tau=\left\{t<0:0;t>T:\chi a T;0<t<T:\chi a t\right\}\). The result (\ref{sol}) is in well concordance with the existing description of the PDC process \cite{walls}.

\section{Particular cases}
In order to establish a link between the model and the physical reality we briefly analyze it for the two relevant particular types of PNES states, namely the twin beam (TWB) and tmo-mode coherently-correlated (TMC) states.

The twin beam states which are also referred to as squeezed vacuum states can be presented in the form (\ref{eq:pnes}) as

\begin{equation}
\label{eq:twb}
| x \rangle\rangle = \sqrt {\left( {1 - x^2} \right)} \sum\limits_n 
{x^n\left| {n,n} \right\rangle\rangle },
\end{equation}
where generally $x\in{\mathbb C}$ and $0\leq |x|\leq 1$, but without loss of generality we assume $x$ as real . 
TWB states are Gaussian and are widely used for CV teleportation and dense coding. 
The marginal statistical distributions of the TWB modes are equal to thermal states.

The TMC states being the degenerate pair-coherent states \cite{agarwal} can be given as

\begin{equation}
\label{eq:tmc}
| \lambda \rangle\rangle =\frac{1}{\sqrt{I_0(2|\lambda|)}}\sum_{n=0}^{\infty}{\frac{\lambda^n}{n!}\left| {n,n} \right\rangle\rangle},
\end{equation}

where $I_0(x)$ is the modified Bessel function. These non-Gaussian states are eigen for the product of the modes' annihilation operators 
$\hat{a_1}\hat{a_2}\left| {\lambda} \right\rangle\rangle=\lambda \left| {\lambda} \right\rangle\rangle$ and possess the sub-Poisson photon-number statistics for each of the modes.

In order to estimate which of the states is closely to be described by the model, while the mode amplitudes have zero mean values 
we introduce the observables which we call pair-quadratures similarly to the mode quadratures and construct them of operators $\hat{A}=\hat{a}_1\hat{a}_2$, $\hat{A}^+=\hat{a}_1^+\hat{a}_2^+$ as $\hat{C_+}=\hat{A}+\hat{A^+}$ and $\hat{C_-}=\frac{\hat{A}-\hat{A^+}}{i}$. We estimate the evolution of a pair-quadrature dispersion $D_C = \mean{\hat{C}^2}-\mean{\hat{C}}^2$, where $C \equiv C_\pm$ within the model directly diffrentiating the dispersion obatined with averaging by the time-dependent solutions:

\begin{equation}
\frac{dD_C}{dt}\Bigg\arrowvert_{model} = \frac{d}{dt}\Big({\dbra{\Psi(t)}\hat{C}^2\dket{\Psi(t)}-\dbra{\Psi(t)}\hat{C}\dket{\Psi(t)}^2}\Big)
\end{equation}

assuming that the state evolution is decribed by the time dependence of state parameters (\ref{sol}). We compare this to the exact evolution of the pair-quadrature dispersion obtained directly by averaging the pair-quadrature derivatives by the particular states either TWB (\ref{eq:twb}) or TMC (\ref{eq:tmc}):

\begin{equation}
\frac{dD_C}{dt}\Bigg\arrowvert_{exact} = \mean{\frac{d\hat{C}}{dt}\hat{C}+\hat{C}\frac{d\hat{C}}{dt}}-2\mean{\hat{C}}\mean{\frac{d\hat{C}}{dt}},
\end{equation}

where $\frac{d\hat{C}}{dt}$ is obtained from the Heisenderg equation $\frac{d\hat{C}}{dt}=\frac{1}{i\hbar}[\hat{C},\hat{H}]$ for the system Hamiltonian (\ref{eq:hamil}).

For the $\hat{C_+}$ pair-quadrature the resulting derivative within the "exact" approach is

\begin{equation}
\frac{dD_{C+}}{dt}\Bigg\arrowvert_{exact} = \frac{2\chi}{\hbar}\mean{\hat{Q}}\mean{\hat{C_+}},
\end{equation}

where $\hat{Q}=\hat{a_0}+\hat{a_0}^+$ is the pumping quadrature. Calculating this derivative for the particular type of states (TWB or TMC) and comparing it to the expression
obtained by calculation through averaging by time-dependent solutions within the model, we get that for TWB

\begin{equation}
\frac{dD_{C+}}{dt}\Bigg\arrowvert_{exact}^{TWB} = \frac{dD_{C+}}{dt}\Bigg\arrowvert_{model}^{TWB}
= 8\chi\alpha\frac{x(x^2+1)}{(1-x^2)^2},
\end{equation}

while for TMC

\begin{equation}
\frac{dD_{C+}}{dt}\Bigg\arrowvert_{exact}^{TMC} = 8\chi\lambda\alpha \neq \frac{dD_{C+}}{dt}\Bigg\arrowvert_{model}^{TMC}
= 4\chi\lambda\alpha,
\end{equation}

thus the built model is more likely to describe the TWB states generation.

\section{Conclusions}
In this work we built the simple model describing the photon-number entangled states generation in the stimulated parametric down conversion process.
At that we didn't make the traditional parametric approximation and considered pumping mode as quantum. The special statistical properties of the states 
enabled obtaining the solution for the motion equations describing the process of generation in terms of the evolution of the sate parameters. The solution represents the relation between time-profiles of the pumping and generated modes. The model is briefly analyzed with respect to the particular types of photon-number entangled states and is more likely to describe the generation of a twin-beam state.

\end{document}